\documentclass[conference]{IEEEtran}
\IEEEoverridecommandlockouts

\usepackage{cite}
\usepackage{amsmath,amssymb,amsfonts}
\usepackage{algorithmic}
\usepackage{graphicx}
\usepackage{textcomp}
\usepackage{balance}
\usepackage{xcolor}
\def\BibTeX{{\rm B\kern-.05em{\sc i\kern-.025em b}\kern-.08em
    T\kern-.1667em\lower.7ex\hbox{E}\kern-.125emX}}
\begin{document}
\title{An Analysis on Rate-Splitting Multiple Access for IRS Aided 6G Communication}

\author{Aditya Jolly$^{\dag}$, Sudip Biswas$^{\dag}$, and Keshav Singh$^{\ddagger}$\\
\IEEEauthorblockA{
$^\dag$Department of ECE, Indian Institute of Information Technology, Guwahati, India.\\
$^\ddagger$Institute of Communications Engineering, National Sun Yat-sen University, Taiwan, R.O.C.\\
Email: aditya.jolly@iiitg.ac.in, sudip.biswas@iiitg.ac.in, keshav.singh@mail.nsysu.edu.tw}
}

\maketitle

\begin{abstract}
Integrating intelligent reflecting surface (IRS) and Rate-Splitting Multiple Access (RSMA) is an effective solution to improve the spectral/energy efficiency in next-generation (beyond 5G (B5G) and 6G) wireless networks. In this paper, we investigate a rate-splitting (RS)-based transmission technique for an IRS-aided communication network involving both near and cell-edge users. In particular, we derive a new architecture called IRS-RS that leverages the interplay between RS and IRS, with an aim to maximize the weighted sum-rate (WSR) of users by selecting the reflecting coefficients at the IRS and designing beamformers at the BS under the constraints of power at the base station (BS), quality of service (QoS) at each user and finite resolution at the IRS. To solve the non-convex WSR maximization problem, we propose an alternating algorithm and compare its performance with baseline non-orthogonal multiple access (NOMA) based transmission for an IRS-aided communication network for both perfect and imperfect CSIT cases. Through numerical results, it is shown that the proposed IRS-RS architecture yields better QoS with respect to the cell-edge users when compared to IRS-NOMA transmission scheme.
\end{abstract}

\begin{IEEEkeywords}
Intelligent reflecting surface (IRS), rate-splitting (RS), non-orthogonal multiple access (NOMA), beamforming.\vspace{-0.5em}
\end{IEEEkeywords}

\section{Introduction}

\IEEEPARstart{W}ith fifth generation (5G) networks being deployed in phases in various parts of the world, the focus has shifted towards developing the next generation of disruptive wireless technologies. In this context of new communication paradigms in the physical layer, intelligent reflecting surface (IRS) assisted communication and rate-splitting multiple access (RSMA) are two technologies that have gained immense popularity. An IRS is a reconfigurable intelligent and software-controlled metasurface consisting of passive elements, wherein each element independently reflects the incident electromagnetic wave after adjusting the phase of each passive element, thereby controlling the wireless propagation environment.
With regards to works on IRS, recently the authors in~\cite{Basar2019} have discussed several latest research trends in the field of IRS-assisted wireless networks. In~\cite{Basar2020}, space shift keying and spatial modulation schemes for IRS were investigated to enhance the network spectral efficiency. 
On a similar note, while a resource allocation scheme for IRS-aided full duplex cognitive radio networks has been studied in~\cite{Xu2020}, beamforming design algorithms for IRS have been studied in~\cite{Zhou2020}--\!\!\cite{Yan2020}. 

On a similar vein, RSMA and non-orthogonal multiple access (NOMA) have also emerged as a remedy for improving the spectrum efficiency of wireless networks. Considering the fact that IRS will play a pivotal role in the next-generation (i.e., 5G$\&$Beyond and 6G) wireless networks as a cost, power and spectrum efficient  technology, recently the authors in~\cite{Zeng2020, Ding2020} have studied the amalgamation of NOMA transmission schemes with IRS to good effect. 
However, while several works including~\cite{Clerckx2016}--\!\!\cite{Yalcin2020} have illustrated the benefits of using RSMA, to the best of the authors' knowledge no work till date has studied the effect of applying RS to an IRS aided network.
For instance, the authors in~\cite{Clerckx2016} have highlighted the recent advancements in rate-splitting (RS) specially for multiple-input and multiple-output (MIMO) networks, while in~\cite{Joudeh2016} the sum rate maximization problem was considered to design precoders with partial channel state information. Further, the authors in~\cite{Joudeh2_2016} studied a RS transmission scheme to achieve max-min fairness in a multi-user multi-input single-output (MISO) system. Similarly, while in \cite{Zhang2020} by considering RS and common beamforming coordination, the joint optimization of beamforming and rate allocation was studied, in~\cite{Yin2020} RSMA was adopted for a multi-group multicast downlink MISO communication system.

%
Based on the above discussion, in this paper, we propose a novel transmission technique based on RS for an IRS-aided communication network involving multiple near and cell-edge users. In particular, we implement RSMA and accordingly select the reflecting coefficients at the IRS and design beamformers at the BS for its broadcasted signal with the objective of maximizing the weighted sum-rate (WSR) of users subject to the constraints of power at the BS, quality of service (QoS) at each user and finite resolution for the elements of the IRS. The proposed RS based algorithm for IRS is then compared with baseline NOMA based transmission for IRS from \cite{Ding2020} in terms of sum-rate for both perfect and imperfect CSIT cases. It is shown that the RS based beamformer design for IRS is a much better transmission technique than NOMA under both perfect and imperfect CSIT, whereby IRS-RS can achieve better rates for the cell-edge users than IRS-NOMA, thus increasing the QoS of network operators and quality of experience (QoE) of users. 

The rest of the paper is structured as follows. Section II gives the network model, while Section III illustrates the design for IRS-RS. The problem formulation is provided in Section IV and the RS based beamformer design is given in Section V. Numerical results are provided in Section VI followed by conclusion in Section VII.
\section{Network Model}\vspace{-0.25em}
\begin{figure}[htbp]
\centerline{\includegraphics[scale=0.15]{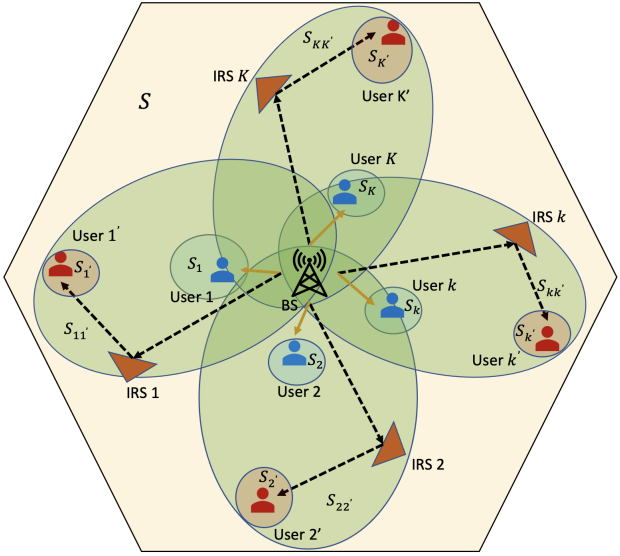}}\vspace{-0.50em}
\caption{An illustration of the network model with $K$ near users, $K$ cell-edge users and $K$ IRS's employing 2-Layer HRS.}\vspace{-0.75em}
\label{fig1}
\end{figure}
\par We consider the downlink communication of an IRS assisted network with $2K$ users being served by a base-station (BS) located at the centre of a hexagonal cell as shown in Fig. \ref{fig1}. Of the $2K$ users in the network, it is assumed that there are $K$ near and $K$ cell-edge users, such that they are indexed as $\mathcal{K}=\{1,\dots, k,\dots, K\}$ and $\mathcal{K}^{'}=\{1^{'},\dots, k^{'},\dots K^{'}\}$, respectively. The IRSs are placed close to the cell-edge users, with each cell-edge user being aided by an IRS and it is assumed that there is no direct link from the BS to the cell-edge user. Thus, there are $K$ IRSs in the network, each having $N$ reflecting surfaces. The BS has $M$ antennas, while each user is equipped with  a single antenna. Let the signal transmitted by the BS in a given channel use be denoted by $\mathbf{x}\in\mathbb{C}^{M\times 1}$. Then, the signal received by the $k$th near user is given as
\begin{align}\label{eq1}
y_{k} = \mathbf{h}^{H}_{k}\mathbf{x} + w_{k},
\end{align}
where $\mathbf{h}_{k}\in\mathbb{C}^{M\times 1}$  is the channel between the BS and the $k$th near user and $w_{k}\in\mathcal{CN}(0,1)$ is the AWGN at the $k$th receiver. 
Similarly, the signal received by the $k^{'}$th cell-edge user from the BS through the $k$th IRS is given as\footnote{This work assumes that there is no direct link between the cell-edge users and the BS and hence each cell-edge user is served by the BS only through the aid of an IRS.}  
\begin{align}\label{eq2}
y_{k^{'}} = \mathbf{h}^{H}_{k^{'}}\boldsymbol{\Theta}_{k}\mathbf{G}_{k}\mathbf{x} + w_{k^{'}},
\end{align}
where $\mathbf{h}_{k^{'}}\in\mathbb{C}^{N\times 1}$ and $\mathbf{G}_{k}\in\mathbb{C}^{N\times M}$ are the channels between $k$th IRS to $k^{'}$th cell-edge user and BS to the $k$th IRS, respectively. $\boldsymbol{\Theta}_{k}\in\mathbb{C}^{N\times N}$ is a diagonal matrix with diagonal entries as $\Theta_{k}(i,i) = \beta_{k,i}e^{-j\theta_{k,i}}$, where, $\beta_{k,i}$ and $\theta_{k,i}$ are the amplitude reflection co-efficient and the reflection phase shift of $k$th IRS's, $i$th element, respectively and $w_{k^{'}}\in\mathcal{CN}(0,1)$ is the AWGN at the $k^{'}$th receiver. In addition, the BS and IRS are integrated through a Central Control Unit (CCU)~\cite{L.Huang2018}. For the sake of tractability, we assume that the BS knows the channel state information (CSI) of each user perfectly with the help of this CCU which is able to estimate all the above mentioned CSIs \cite{L.Huang2018}. Nevertheless, the scenario of imperfect CSIT will be implemented for the proposed algorithm in the numerical results. 

\section{Design of IRS-RS}
In this section, we apply a 2-Layer Hierarchal RS (HRS) architecture to the IRS aided network\footnote{We choose a 2-layer HRS architecture as it is a more general approach (i.e., 1-layer HRS is a sub-scheme of 2-layer HRS) and it ensures the architecture's scalability to different cases.}.
\subsubsection{Architecture for RSMA}The 2$K$ users in the network are divided into $K$ groups as: $\{ (1,1^{'}),(2,2^{'}),\dots,(k,k^{'}),\dots,(K,K^{'})\}$. 
The message of each user, near or cell-edge is split into three parts and encoded into
\begin{align}
\mathbf{m}_{k} = [M_{k1},M_{k2},M_{k3}]^{T}
\end{align}
for near users, with $k\in\mathcal{K}$, and 
\begin{align}
\mathbf{m}_{k^{'}} = [M_{k^{'}1},M_{k^{'}2},M_{k^{'}3}]^{T}
\end{align}
for cell-edge users, with $k^{'}\in\mathcal{K}^{'}$. The common stream for all the groups is
\begin{align}
s =  \sum\limits_{k=1}^K M_{k1} +\sum\limits_{k'=1}^{K'} M_{k^{'}1},
\end{align}
with the common stream for the $k^{th}$ group given as 
\begin{align}
s_{kk^{'}} =  M_{k2} + M_{k^{'}2}. 
\end{align}
The private streams are $s_{k}= M_{k3}$ for the near user $k$ and $s_{k^{'}}= M_{k^{'}3}$ for the cell-edge user $k^{'}$. 

We now consider the following beamforming matrix that will be used by the BS to precode its transmitted signal:
\begin{align}
\mathbf{P} &= [\mathbf{p},\mathbf{p}_{11^{'}},\dots,\mathbf{p}_{KK^{'}},\mathbf{p}_{1},\dots,\mathbf{p}_{K},\mathbf{p}_{1^{'}},\dots,\mathbf{p}_{K^{'}}].
\end{align}
Here, $\mathbf{p}\in\mathbb{C}^{M\times1}$ is the precoder for the common stream for all the groups, $\mathbf{p}_{kk'}\in\mathbb{C}^{M\times1}$ is the precoder for the common stream of the group $(k,k')$, $\mathbf{p}_{k}\in\mathbb{C}^{M\times1}$ is the precoder for the private stream of the near user $k$ and $\mathbf{p}_{k^{'}}\in\mathbb{C}^{M\times1}$ is the precoder for the private stream of the cell-edge user $k^{'}$.
Now applying RS, the overall data stream to be transmitted by the BS is given as
\begin{align}
\mathbf{s}=[s, s_{kk^{'}}, s_k, s_{k^{'}}]^T.
\end{align}
Accordingly, the resulting RS signal broadcasted by the BS is 
\begin{align}
\textbf{x} &=\mathbf{Ps}\nonumber\\
&=\mathbf{p}s + \sum\limits_{k} \mathbf{p}_{kk^{'}}s_{kk^{'}} + \sum\limits_{k} (\mathbf{p}_{k}s_{k} + \mathbf{p}_{k^{'}}s_{k^{'}}).
\end{align}
We assume that $\text{tr}(\mathbf{m}_{k}\mathbf{m}_{k}^{H}) = \textbf{I}$, $\text{tr}(\mathbf{m}_{k^{'}}\mathbf{m}_{k^{'}}^{H}) = \textbf{I}$ and the total transmit power is constrained by $\text{tr}(\mathbf{P}\mathbf{P}^{H}) \leq P_{t}$. Since the noise variances at the receivers are considered to be one for all near and cell-edge users as seen in \eqref{eq1} and \eqref{eq2}, the transmit SNR is equivalent to the total power consumption $P_t$.

\subsubsection{Finite resolution beamforming for IRS} The cell-edge users' effective channel vectors are determined by the choice of $\boldsymbol{\Theta}$. In literature \cite{Basar2019, Basar2020, Xu2020}, the choices for $\beta_{k,i}$ and $\theta_{k,i}$ are usually  arbitrary. However, due to the limitations on hardware of the IRS this assumption may not always hold true. Hence, in order to obtain insights on the novel problem of applying RS in an IRS aided network, in this paper we assume finite resolution beamforming through the ON-OFF keying approach, whereby each diagonal element of $\boldsymbol{\Theta}$ is either 1 (on) or 0 (off), like the one proposed in \cite{Ding2020}. In particular, $\boldsymbol{\Theta}$ is designed as $\mathbf{Z} = \text{diag}\{\mathbf{a}_{p}\}$, where $\mathbf{a}_{p}$ is the $p$th column of $\mathbf{A}$, which is subject to the constraints $\mathbf{a}_{p}^{H}\mathbf{a}_{z} = 0$ for $p \neq z$, and $\mathbf{a}_{p}^{H}\mathbf{a}_{p} = 1$. Here, $\mathbf{A} = \frac{1}{\sqrt{Q}}\mathbf{I}_{P} \otimes \mathbf{1}_{Q}$, with $\mathbf{I}_{P}$ being a $P \times P$ identity matrix and $\mathbf{1}_{Q}$ is a $Q \times 1$ all-ones vector. Further, $P$ and $Q$ are integers, such that the number of IRS elements $N = PQ$.

\section{Problem Formulation}
In this section we propose the RSMA based precoder design problem for the IRS aided network with the objective of maximizing the weighted sum-rate (WSR) of users subject to the constraints of power at the BS and quality of service (QoS) at each user. 
We start by performing successive interference cancellation (SIC) at user $K^{'}$ and obtaining the SINR equations as \eqref{SINR1}, \eqref{SINR2} and \eqref{SINR3}. Similarly, performing SIC at the near user $k$ we obtain the SINR equations \eqref{SINR4}, \eqref{SINR5} and \eqref{SINR6}.
\begin{figure*}[!t]
\normalsize
\setcounter{equation}{9}
\begin{equation}\label{SINR1}
{\Gamma}_{s}^{k^{'}} = \frac{\|\mathbf{h}_{k^{'}}^{H}\mathbf{Z}_{k}\mathbf{G}_{k}\mathbf{p}\|^{2}} {\sum\limits_{k}(\|\mathbf{h}_{k^{'}}^{H}\mathbf{Z}_{k}\mathbf{G}_{k}\mathbf{p}_{kk^{'}} \|^{2} + \|\mathbf{h}_{k^{'}}^{H}\mathbf{Z}_{k}\mathbf{G}_{k}\mathbf{p}_{k}\|^{2} + \|\mathbf{h}_{k^{'}}^{H}\mathbf{Z}_{k}\mathbf{G}_{k}\mathbf{p}_{k^{'}}\|^{2}) +1}
\end{equation}
\begin{equation}\label{SINR2}
{\Gamma}_{s_{kk^{'}}}^{k^{'}} = \frac{\|\mathbf{h}_{k^{'}}^{H}\mathbf{Z}_{k}\mathbf{G}_{k}\mathbf{p}_{kk^{'}}\|^{2}}{\sum\limits_{i\neq k}\|\mathbf{h}_{i^{'}}^{H}\mathbf{Z}_{i}\mathbf{G}_{i}\mathbf{p}_{ii^{'}} \|^{2} +\sum\limits_{k}( \|\mathbf{h}_{k^{'}}^{H}\mathbf{Z}_{k}\mathbf{G}_{k}\mathbf{p}_{k}\|^{2} + \|\mathbf{h}_{k^{'}}^{H}\mathbf{Z}_{k}\mathbf{G}_{k}\mathbf{p}_{k^{'}}\|^{2}) + 1}
\end{equation}
\begin{equation}\label{SINR3}
{\Gamma}_{s_{k^{'}}} = \frac{\|\mathbf{h}_{k^{'}}^{H}\mathbf{Z}_{k}\mathbf{G}_{k}\mathbf{p}_{k^{'}}\|^{2}}{\|\mathbf{h}_{k^{'}}^{H}\mathbf{Z}_{k}\mathbf{G}_{k}\mathbf{p}_{k}\|^{2} +\sum\limits_{i\neq k}( \|\mathbf{h}_{i^{'}}^{H}\mathbf{Z}_{i}\mathbf{G}_{i}\mathbf{p}_{i}\|^{2} + \|\mathbf{h}_{i^{'}}^{H}\mathbf{Z}_{i}\mathbf{G}_{i}\mathbf{p}_{i^{'}}\|^{2}+ \|\mathbf{h}_{i^{'}}^{H}\mathbf{Z}_{i}\mathbf{G}_{i}\mathbf{p}_{ii^{'}} \|^{2})+1}
\end{equation}
\begin{equation}\label{SINR4}
{\Gamma}_{s}^{k} = \frac{\|\mathbf{h}_{k}^{H}\mathbf{p}\|^{2}} {\sum\limits_{k}(\|\mathbf{h}_{k}^{H}\mathbf{p}_{kk^{'}} \|^{2} + \|\mathbf{h}_{k}^{H}\mathbf{p}_{k}\|^{2} + \|\mathbf{h}_{k}^{H}\mathbf{p}_{k^{'}}\|^{2}) + 1}
\end{equation}
\begin{equation}\label{SINR5}
 {\Gamma}_{s_{kk^{'}}}^{k} = \frac{\|\mathbf{h}_{k}^{H}\mathbf{p}_{kk^{'}}\|^{2}}{\sum\limits_{i\neq k}\|\mathbf{h}_{i}^{H}\mathbf{p}_{ii^{'}} \|^{2} +\sum\limits_{k}( \|\mathbf{h}_{k}^{H}\mathbf{p}_{k}\|^{2} + \|\mathbf{h}_{k}^{H}\mathbf{p}_{k^{'}}\|^{2}) + 1}
\end{equation}
\begin{equation}\label{SINR6}
{\Gamma}_{s_{k}} = \frac{\|\mathbf{h}_{k}^{H}\mathbf{p}_{k}\|^{2}}{\|\mathbf{h}_{k}^{H}\mathbf{p}_{k^{'}}\|^{2} +\sum\limits_{i\neq k}( \|\mathbf{h}_{i}^{H}\mathbf{p}_{i}\|^{2} + \|\mathbf{h}_{i}^{H}\mathbf{p}_{i^{'}}\|^{2}+\|\mathbf{h}_{i}^{H}\mathbf{p}_{ii^{'}} \|^{2})+1}
\end{equation}
\hrulefill
\end{figure*}
The corresponding rates for the $k$ and the $k^{'}$ users are respectively given as
\begin{align} 
&R_{s}^{k} = \log_{2}(1 + {\Gamma}_{s}^{k}), \\
&R_{s_{kk^{'}}}^{k} = \log_{2}(1 + {\Gamma}_{s_{kk^{'}}}^{k}), \\
&R_{s_{k}} = \log_{2}(1 + {\Gamma}_{s_{k}}), \\
&R_{s}^{k^{'}} = \log_{2}(1 + {\Gamma}_{s}^{k^{'}}), \\
&R_{s_{kk^{'}}}^{k^{'}} = \log_{2}(1 + {\Gamma}_{s_{kk^{'}}}^{k^{'}}), \\
&R_{s_{k^{'}}} = \log_{2}(1 + {\Gamma}_{s_{k^{'}}}).
\end{align}
The common rate per user is given as
\begin{align}
R_{s} = \min{\{R_{s}^{1},R_{s}^{2},R_{s}^{3},..,R_{s}^{k},R_{s}^{1^{'}},R_{s}^{2^{'}},R_{s}^{3^{'}},\dots,R_{s}^{k^{'}}\}}.
\end{align}
 The common rate per group is given as
 \begin{align}
 R_{s_{kk^{'}}} = \min{\{R_{s_{kk^{'}}}^{k},R_{s_{kk^{'}}}^{k^{'}}\}}.
 \end{align}
$R_{s}$ is shared among users such that $C_{k}^{s}$ is the $k$th user's portion and $C_{k^{'}}^{s}$   is the $k^{'}$th  user's portion of the common rate such that
\begin{align}
 \sum\nolimits_{i\in{\mathcal{K}}}\left(C_{i}^{s}+C_{i^{'}}^{s}\right) = R_{s}. 
 \end{align}
$R_{s_{kk^{'}}}$ is shared among the group users $k$ and $k^{'}$ such that $C_{k}^{s_{kk^{'}}}$ and $C_{k^{'}}^{s_{kk^{'}}}$ are the $k$th and $k^{'}$th users portion of the common rate in the group, respectively such that
 \begin{align}
 C_{k}^{s_{kk^{'}}}+C_{k^{'}}^{s_{kk^{'}}}= R_{s_{kk^{'}}}. 
 \end{align}
 The total rate for user $k$ and $k^{'}$, respectively are given as
 \begin{align}
 R_{k,\text{tot}} = C_{k}^{s} + C_{k}^{s_{kk^{'}}} + R_{s_{k}},
 \end{align}
 and 
 \begin{align}
 R_{k^{'}\!,\text{tot}} = C_{k^{'}}^{s} + C_{k^{'}}^{s_{kk^{'}}} + R_{s_{k^{'}}}.
 \end{align}

Now with the derived rates, we can form the WSR Problem for the finite resolution IRS aided network. For a given weight vector $\textbf{u} = [u_{1},\dots,u_{k},\dots, u_{K},u_{1^{'}}, \dots,u_{k^{'}},\dots, u_{K^{'}}]$, the WSR problem for the 2-layer HRS is given as \vspace{-0.5em}
 \begin{align}
 R(\mathbf{u}) = &\max_{\mathbf{P}, \, \mathbf{c},\, \mathbf{a}_{p} }\sum_{i\in\mathcal{K}\,,i^{'}\in\mathcal{K}^{'}}\left(u_{i}R_{i,\text{tot}} + u_{i^{'}}R_{i^{'},\text{tot}}\right) \label{e31} \\
\text{subject to}  \;\;
  &\text{C1:}\quad \sum\nolimits_{i\in\mathcal{K}\,,i^{'}\in\mathcal{K}^{'}}(C_{i}^{s}+C_{i^{'}}^{s}) \leq R_{s}\nonumber\\
      &  \text{C2:}\quad C_{k}^{s_{kk^{'}}}+C_{k^{'}}^{s_{kk^{'}}} \leq R_{s_{kk^{'}}}, k\in\mathcal{K}, k^{'}\in\mathcal{K}^{'}\nonumber\\
      &  \text{C3:}\quad \text{tr}(\mathbf{P}\mathbf{P}^{H}) \leq P_{t}\nonumber\\
       & \text{C4:}\quad R_{k,tot} \geq R^{th}_{k}, k\in\mathcal{K}\nonumber\\
       & \text{C5:}\quad R_{k^{'},tot} \geq R^{th}_{k^{'}}, k^{'}\in\mathcal{K}^{'}\nonumber\\
       & \text{C6:}\quad \mathbf{c} \geq 0,\nonumber\\
       & \text{C7:}\quad \mathbf{a}_{p}^H \mathbf{a}_{z}= \begin{cases}
      1 & \text{if $p= z$}\\
      0 & \text{otherwise.}
    \end{cases}   \nonumber
  \end{align}
where $\mathbf{c}$ is the common rate vector formed by $[C_{k}^{s},C_{k}^{s_{kk^{'}}}] \, \forall  \{k \in \mathcal{K}, k^{'}\in\mathcal{K}^{'}\}$. In its current form, the problem \eqref{e31} is difficult to solve. However, for a given weight vector, the problem can be solved by applying a modified WMMSE approach, which is derived in the following section. 
\section{RSMA based Beamformer Design}
We first consider the cell-edge user-$1^{'}$ for illustration. This procedure is the same for all the users and can be extended easily to the set $\mathcal{K}$ and $\mathcal{K}^{'}$, whereby while we work with \eqref{eq1} for $\mathcal{K}$, we must work with \eqref{eq2} for $\mathcal{K}^{'}$. Accordingly, the signal received at user-$1^{'}$ is $y_{1^{'}} = \mathbf{h}^{H}_{1^{'}}\mathbf{Z}_{1}\mathbf{G}_{1}\mathbf{x} + w_{1^{'}}$. It decodes three streams, $s$ (the common stream for all), $s_{11^{'}}$ (the common stream per group) and $s_{1^{'}}$ (user $1^{'}$'s private stream) sequentially using successive interference cancellation (SIC) at the receiver. $s$ is decoded first and estimated as $\hat{s} = g^{s}_{1^{'}}y_{1^{'}}$, where $g^{s}_{1^{'}}$ is the equaliser. After decoding and removing $s$ from $y_{1^{'}}$, the estimate of the common stream per group $\hat{s}_{11^{'}} = g^{s_{11^{'}}}_{1^{'}}(y_{1^{'}} - \mathbf{h}^{H}_{1^{'}}\mathbf{Z}_{1}\mathbf{G}_{1}\mathbf{p}s )$, and finally the estimate of the private stream is obtained after removing the per group common component from the signal, $\hat{s}_{1^{'}}= g^{s_{1^{'}}}_{1^{'}}(y_{1^{'}} - \mathbf{h}^{H}_{1^{'}}\mathbf{Z}_{1}\mathbf{G}_{1}\mathbf{p}s - \mathbf{h}^{H}_{1^{'}}\mathbf{Z}_{1}\mathbf{G}_{1}\mathbf{p}_{11^{'}}s_{11^{'}}) $. Here, $g^{s}_{1^{'}}$, $g^{s_{11^{'}}}_{1^{'}}$ and $g^{s_{1^{'}}}_{1^{'}}$ are the equalisers for the user-$1^{'}$. 

When MMSE receiver is used, the SINR-based beamformers are equivalent to MSE-based ones as they are related to each other as $\text{MSE}=1/(1+\text{SINR})$. Hence, rate-based beamformer design problems utilizing $\log_2(1 + \text{SINR})$ can be converted into MSE-based ones as $-\log_2(\text{MSE})$. Since, MSE-based designs are easier to manipulate, hereinafter we consider the minimization of MSE of a user to relate to the maximization of a tight lower bound of its rate.
Accordingly, by defining the MSE of each stream as $\varepsilon_{k^{'}} \triangleq \mathbb{E}\{ | s_{k^{'}} - \hat{s}_{k^{'}}|\} $, the MSEs for the network are calculated as
\begin{align}\label{MSE1}
\varepsilon^{s}_{1^{'}} &= |g^{s}_{1^{'}} | T^{s}_{1^{'}}  - 2Re\{ g^{s}_{1^{'}}\mathbf{h}^{H}_{1^{'}}\mathbf{Z}_{1}\mathbf{G}_{1}\mathbf{p}\} + 1,
\end{align}
\begin{align}\label{MSE2}
\varepsilon^{s_{11^{'}}}_{1^{'}} &= | g^{s_{11^{'}}}_{1^{'}} | T^{s_{11^{'}}}_{1^{'}}  - 2Re\{ g^{s_{11^{'}}}_{1^{'}}\mathbf{h}^{H}_{1^{'}}\mathbf{Z}_{1}\mathbf{G}_{1}\mathbf{p}_{11^{'}}\} + 1,
\end{align}
\begin{align}\label{MSE3}
\varepsilon^{s_{1^{'}}}_{1^{'}} &= | g^{s_{1^{'}}}_{1^{'}} | T^{s_{1^{'}}}_{1^{'}}  - 2Re\{ g^{s_{1^{'}}}_{1^{'}}\mathbf{h}^{H}_{1^{'}}\mathbf{Z}_{1}\mathbf{G}_{1}\mathbf{p}_{1^{'}}\} + 1.
\end{align}
Here, $T^{s}_{1^{'}} \triangleq \|\mathbf{h}^{H}_{1^{'}}\mathbf{Z}_{1}\mathbf{G}_{1}\mathbf{p}\|^{2} + \|\mathbf{h}^{H}_{1^{'}}\mathbf{Z}_{1}\mathbf{G}_{1}\mathbf{p}_{11^{'}}\|^{2} + \|\mathbf{h}^{H}_{1^{'}}\mathbf{Z}_{1}\mathbf{G}_{1}\mathbf{p}_{1}\|^{2} + \|\mathbf{h}^{H}_{1^{'}}\mathbf{Z}_{1}\mathbf{G}_{1}\mathbf{p}_{1^{'}}\|^{2} + 1$ is the receive power at user-$1^{'}$, $T^{s_{11^{'}}}_{1^{'}} \triangleq T^{s}_{1^{'}} - \|\mathbf{h}^{H}_{1^{'}}\mathbf{Z}_{1}\mathbf{G}_{1}\mathbf{p}\|^{2}$, and $T^{s_{1^{'}}}_{1^{'}} \triangleq T^{s_{11^{'}}}_{1^{'}} - \|\mathbf{h}^{H}_{1^{'}}\mathbf{Z}_{1}\mathbf{G}_{1}\mathbf{p}_{11^{'}}\|^{2}$. 

Now, differentiating the MSE terms in \eqref{MSE1}--\eqref{MSE3} with respect to $g^{s}_{1^{'}}$, $g^{s_{11^{'}}}_{1^{'}}$, and $g^{s_{1^{'}}}_{1^{'}}$ and equating them to zero, the optimum MMSE equalizers can be obtained as
\begin{align}
(g^{s}_{1^{'}})^{\text{MMSE}} &= (\mathbf{p})^{H}(\mathbf{h}^{H}_{1^{'}}\mathbf{Z}_{1}\mathbf{G}_{1})^{H}(T^{s}_{1^{'}})^{-1},\label{MMSE1}
\\
(g^{s_{11^{'}}}_{1^{'}})^{\text{MMSE}} &= (\mathbf{p}_{11^{'}})^{H}(\mathbf{h}^{H}_{1^{'}}\mathbf{Z}_{1}\mathbf{G}_{1})^{H}( T^{s_{11^{'}}}_{1^{'}})^{-1},\label{MMSE2}
\\
(g^{s_{1^{'}}}_{1^{'}} )^{\text{MMSE}} &= (\mathbf{p}_{1^{'}})^{H}(\mathbf{h}^{H}_{1^{'}}\mathbf{Z}_{1}\mathbf{G}_{1})^{H}( T^{s_{1^{'}}}_{1^{'}})^{-1}.\label{MMSE3}
\end{align}
Substituting \eqref{MMSE1}--\eqref{MMSE3} in \eqref{MSE1}--\eqref{MSE3}, the MMSEs become
\begin{align}
(\varepsilon^{s}_{1^{'}})^{\text{MMSE}}& \triangleq \min_{g^{s}_{1^{'}}} \varepsilon^{s}_{1^{'}} = (T^{s}_{1^{'}})^{-1} I^{s}_{1^{'}},\label{MMSE4}
\\
(\varepsilon^{s_{11^{'}}}_{1^{'}})^{\text{MMSE}} &\triangleq \min_{g^{s_{11^{'}}}_{1^{'}}} \varepsilon^{s_{11^{'}}}_{1^{'}} = (T^{s_{11^{'}}}_{1^{'}})^{-1} I^{s_{11^{'}}}_{1^{'}},\label{MMSE5}
\\
(\varepsilon^{s_{1^{'}}}_{1^{'}} )^{\text{MMSE}}& \triangleq \min_{g^{s_{1^{'}}}_{1^{'}}} \varepsilon^{s_{1^{'}}}_{1^{'}}  = (T^{s_{1^{'}}}_{1^{'}})^{-1} I^{s_{1^{'}}}_{1^{'}}.\label{MMSE6}
\end{align}
Here, $I^{s}_{1^{'}} = T^{s_{11^{'}}}_{1^{'}}$,  $I^{s_{11^{'}}}_{1^{'}} = T^{s_{1^{'}}}_{1^{'}}$ and $I^{s_{1^{'}}}_{1^{'}} = T^{s_{1^{'}}}_{1^{'}} - \|\mathbf{h}^{H}_{1^{'}}\mathbf{Z}_{1^{'}}\mathbf{G}_{1^{'}}\mathbf{p}_{1^{'}}\|^{2}$. Based on \eqref{MMSE4}--\eqref{MMSE6}, the SINRs for decoding the intended streams at user-$1^{'}$ can be expressed as ${\Gamma}_{S}^{1^{'}} = \frac{1}{(\varepsilon^{s}_{1^{'}})^{\text{MMSE}}} - 1$, ${\Gamma}_{s_{11^{'}}}^{1^{'}} = \frac{1}{(\varepsilon^{s_{11^{'}}}_{1^{'}})^{\text{MMSE}}} - 1$ and ${\Gamma}_{s_{1^{'}}} = \frac{1}{(\varepsilon^{s_{1^{'}}}_{1^{'}} )^{\text{MMSE}}} - 1$. 

Now based on the equivalence of MSE and rate as discussed earlier, the equivalent rates in the network are given as $R_{s}^{1^{'}} = -\log((\varepsilon^{s}_{1^{'}})^{\text{MMSE}})$, $R_{s_{11^{'}}}^{1^{'}} = -\log((\varepsilon^{s_{11^{'}}}_{1^{'}})^{\text{MMSE}})$ and $R_{s_{1^{'}}} = -\log((\varepsilon^{s_{1^{'}}}_{1^{'}} )^{\text{MMSE}})$, whereby the corresponding WMSEs are given as
\begin{align}
\xi^{s}_{1^{'}} &= u^{s}_{1^{'}}\varepsilon^{s}_{1^{'}} - \log(u^{s}_{1^{'}}),\label{MMSE7}\\
\xi^{s_{11^{'}}}_{1^{'}} &= u^{s_{11^{'}}}_{1^{'}}\varepsilon^{s_{11^{'}}}_{1^{'}} - \log(u^{s_{11^{'}}}_{1^{'}}),\label{MMSE8}\\
\xi^{s_{1^{'}}}_{1^{'}} &= u^{s_{1^{'}}}_{1^{'}}\varepsilon^{s_{1^{'}}}_{1^{'}} - \log(u^{s_{1^{'}}}_{1^{'}}).\label{MMSE9}
\end{align}
Here, $u^{s}_{1^{'}}$, $u^{s_{11^{'}}}_{1^{'}}$ and $u^{s_{1^{'}}}_{1^{'}}$ are weights associated with each stream at user-$1^{'}$. By differentiating \eqref{MMSE7}--\eqref{MMSE9} with respect to $g^{s}_{1^{'}}$, $g^{s_{11^{'}}}_{1^{'}}$, and $g^{s_{1^{'}}}_{1^{'}}$ and equating them to zero,  we derive the optimum equalisers $(g^{s}_{1^{'}})^{*} = (g^{s}_{1^{'}})^{\text{MMSE}}$,  $(g^{s_{11^{'}}}_{1^{'}})^{*} = (g^{s_{11^{'}}}_{1^{'}})^{\text{MMSE}}$ and  $(g^{s_{1^{'}}}_{1^{'}})^{*} = (g^{s_{1^{'}}}_{1^{'}})^{\text{MMSE}}$. Now, by substituting these optimum equalisers in \eqref{MMSE7}--\eqref{MMSE9} we obtain
\begin{align}
\xi^{s}_{1^{'}}((g^{s}_{1^{'}})^{\text{MMSE}}) &= u^{s}_{1^{'}}(\varepsilon^{s}_{1^{'}})^{\text{MMSE}} - \log(u^{s}_{1^{'}}), \label{MMSE10}\\
\xi^{s_{11^{'}}}_{1^{'}} ((g^{s_{11^{'}}}_{1^{'}})^{\text{MMSE}})&= u^{s_{11^{'}}}_{1^{'}}(\varepsilon^{s_{11^{'}}}_{1^{'}})^{\text{MMSE}} - \log(u^{s_{11^{'}}}_{1^{'}}), \label{MMSE11}\\
\xi^{s_{1^{'}}}_{1^{'}}((g^{s_{1^{'}}}_{1^{'}})^{\text{MMSE}}) &= u^{s_{1^{'}}}_{1^{'}}(\varepsilon^{s_{1^{'}}}_{1^{'}})^{\text{MMSE}} - \log(u^{s_{1^{'}}}_{1^{'}}). \label{MMSE12}
\end{align}
By further differentiating \eqref{MMSE10}--\eqref{MMSE12} with respect to $u^{s}_{1^{'}}$, $u^{s_{11^{'}}}_{1^{'}}$ and $u^{s_{1^{'}}}_{1^{'}}$ and equating them to zero we obtain the optimum MMSE weights as 
\begin{align}
(u^{s}_{1^{'}})^{*} = &(u^{s}_{1^{'}})^{\text{MMSE}} &\triangleq ((\varepsilon^{s}_{1^{'}})^{\text{MMSE}})^{-1},  \label{MMSE13}
\\
(u^{s_{11^{'}}}_{1^{'}})^{*} = &(u^{s_{11^{'}}}_{1^{'}})^{\text{MMSE}} &\triangleq ((\varepsilon^{s_{11^{'}}}_{1^{'}})^{\text{MMSE}})^{-1}, 
\label{MMSE14}\\
(u^{s_{1^{'}}}_{1^{'}})^{*} =& (u^{s_{1^{'}}}_{1^{'}})^{\text{MMSE}} &\triangleq (\varepsilon^{s_{1^{'}}}_{1^{'}})^{\text{MMSE}})^{-1}. \label{MMSE15}
\end{align}
\par Now, substituting \eqref{MMSE13}--\eqref{MMSE15} in \eqref{MMSE10}--\eqref{MMSE12}, respectively we establish the relationship between WMMSE and rate as
\begin{align}
(\xi^{s}_{1^{'}})^{\text{MMSE}} \triangleq \min_{u^{s}_{1^{'}}, g^{s}_{1^{'}}}\xi^{s}_{1^{'}} = 1 - R_{s}^{1^{'}},
\end{align}
\begin{align}
(\xi^{s_{11^{'}}}_{1^{'}})^{\text{MMSE}} \triangleq \min_{u^{s_{11^{'}}}_{1^{'}}, g^{s_{11^{'}}}_{1^{'}}}\xi^{s_{11^{'}}}_{1^{'}} = 1 - R_{s_{11^{'}}}^{1^{'}},
\end{align}
\begin{align}
(\xi^{s_{1^{'}}}_{1^{'}})^{\text{MMSE}} \triangleq \min_{u^{s_{1^{'}}}_{1^{'}}, g^{s_{1^{'}}}_{1^{'}}}
\xi^{s_{1^{'}}}_{1^{'}} = 1 - R_{s_{1^{'}}}^{1^{'}}
\end{align}
\par Similarly, we can not only establish the WMMSE-rate relationships for near user-1 with the essential difference being the channel matrix, but also for all the other users in the set $\{\mathcal{K}\cup\mathcal{K}^{'}\}$. Based on the derived WMMSE-rate relationship, we now reformulate the optimisation problem for IRS-RS as
\begin{align}
 R_{\text{WMMSE}} \simeq &\min_{\mathbf{P}, \, \mathbf{d}, \, \mathbf{u}, \, \mathbf{g}, \, \mathbf{a}_{p} }\sum_{i\in\mathcal{K}\,,i^{'}\in\mathcal{K}^{'}}\left(u_{i}\xi_{i,\text{tot}} + u_{i^{'}}\xi_{i^{'},\text{tot}}\right) \label{eWMMSE} \\
\text{subject to} \;\;
  &\text{C1:}\quad \sum\nolimits_{i\in\mathcal{K}\,,i^{'}\in\mathcal{K}^{'}}(D^{s}_{i} + D^{s}_{i^{'}}) + 1 \geq \xi_{s}\nonumber\\
      &  \text{C2:}\quad D^{s_{kk^{'}}}_{k} + D^{s_{kk^{'}}}_{k^{'}} + 1 \geq \xi_{s_{kk^{'}}}, k\in\mathcal{K}, k^{'}\in\mathcal{K}^{'}\nonumber\\
      &  \text{C3:}\quad \text{tr}(\mathbf{P}\mathbf{P}^{H}) \leq P_{t}\nonumber\\
       & \text{C4:}\quad \xi_{k,tot} \leq 1-R^{th}_{k}, k\in\mathcal{K}\nonumber\\
       & \text{C5:}\quad \xi_{k^{'},tot} \leq 1-R^{th}_{k^{'}}, k^{'}\in\mathcal{K}^{'}\nonumber\\
       & \text{C6:}\quad \mathbf{d} \leq 0,\nonumber\\
         & \text{C7:}\quad \mathbf{a}_{p}^H \mathbf{a}_{z}= \begin{cases}
      1 & \text{if $p= z$}\\
      0 & \text{otherwise.}
    \end{cases}   \nonumber
\end{align}
where 
$\textbf{d} = [D_{1}^{s}, D_{2}^{s},\dots, D_{K}^{s}, D_{1^{'}}^{s}, D_{2^{'}}^{s},\dots , D_{K^{'}}^{s}, D_{1}^{s_{11^{'}}},\dots, \\D_{K}^{s_{KK^{'}}}
D_{1^{'}}^{s_{11^{'}}},\dots, D_{K^{'}}^{s_{KK^{'}}}, D_{1}^{s_{1}},\dots, D_{K}^{s_{K}}, D_{1^{'}}^{s_{1^{'}}},\dots, D_{K^{'}}^{s_{K^{'}}} ]$
 is the transformation of the common rate $\mathbf{c}$,
  $\mathbf{u} = [u_{1}^{s}, u_{2}^{s},\dots,u_{K}^{s}, u_{1^{'}}^{s}, u_{2^{'}}^{s},\dots, u_{K^{'}}^{s}, u_{1}^{s_{11^{'}}},\dots, u_{K}^{s_{KK^{'}}},
 u_{1^{'}}^{s_{11^{'}}},\\ \dots,u_{K^{'}}^{s_{KK^{'}}}, u_{1}^{s_{1}},\dots, u_{K}^{s_{K}}, u_{1^{'}}^{s_{1^{'}}},\dots, u_{K^{'}}^{s_{K^{'}}} ]$, $\mathbf{g} = [g_{1}^{s}, g_{2}^{s},.., g_{K}^{s}, g_{1^{'}}^{s}, g_{2^{'}}^{s}, .. , g_{K^{'}}^{s}, g_{1}^{s_{11^{'}}}, \dots , g_{K}^{s_{KK^{'}}},
 g_{1^{'}}^{s_{11^{'}}}, \dots, g_{K^{'}}^{s_{KK^{'}}}\\ g_{1}^{s_{1}},\dots, g_{K}^{s_{K}}, g_{1^{'}}^{s_{1^{'}}},\dots, g_{K^{'}}^{s_{K^{'}}} ]$, $\xi_{k,tot} = D_{k}^{s} + D_{k}^{s_{kk^{'}}} + D_{k}^{s_{k}}$ and $\xi_{k^{'},tot} = D_{k^{'}}^{s} + D_{k^{'}}^{s_{kk^{'}}} + D_{k^{'}}^{s_{k^{'}}}$ are the individual WMSEs for the $k$th near and $k^{'}$th cell-edge user, respectively. Further, $\xi_{s} = \max \{ \xi^{s}_{1}, \xi^{s}_{2},\dots, \xi^{s}_{K}, \xi^{s}_{1^{'}}, \xi^{s}_{2^{'}},\dots, \xi^{s}_{K^{'}} \}$ and $\xi_{s_{kk^{'}}} = \max \{ \xi^{s_{kk^{'}}}_{k}, \xi^{s_{kk^{'}}}_{k^{'}} \}$ are the achievable WMSEs of the overall common stream and the per group common stream, respectively.
 
The Problem \eqref{eWMMSE} is not jointly convex over all the optimization variables. However, it is component wise convex over $\mathbf{P}$, $\mathbf{d}$ and $\mathbf{a}_{p}$ once the other variables are fixed.  When $(\mathbf{P}, \mathbf{d}, \mathbf{u}, \mathbf{a}_{p})$ are fixed, the optimal equalizer is the MMSE equalizer $\mathbf{g}^{\text{MMSE}}$. When $(\mathbf{P,} \mathbf{d}, \mathbf{g}, \mathbf{a}_{p})$ are fixed, the optimal weight is the MMSE weight $\mathbf{u}^{\text{MMSE}}$. Now, we formulate an alternating optimization algorithm as shown in Table \ref{table1}, where in each iteration the solution to \eqref{eWMMSE} is calculated, as a convex optimization problem, assuming an alternatively fixed $\mathbf{P}$ or $\mathbf{d}$. The iterations for optimization continue until convergence or a fixed number of iterations is reached.  
%
    \begin{table}[t]
    			 	 		\renewcommand{\arraystretch}{1.25}
    			 	 		\caption{Alternating optimization algorithm}
    			 	 		\label{table1}
    			 	 		\centering
    			 	 		\begin{tabular}{p{8.25cm}}
    			 	 			\hline
    			 	 			1) Set the iteration number $n=0$, define $\mathbf{a}_{p}$, and initialize $\mathbf{P}^{[n]}$\\
    2) $n\leftarrow n+1$. Select $\mathbf{a}_{p}^{[n]}$ by solving \eqref{eWMMSE} under fixed $\mathbf{P}^{[n-1]}$, $\mathbf{d}^{[n-1]}$\\
    3) Update $\mathbf{d}^{[n]}$ by solving \eqref{eWMMSE} under fixed $\mathbf{P}^{[n-1]}$,  $\mathbf{a}_{p}^{[n-1]}$\\
    4) Update $\mathbf{g}$ and $\mathbf{u}$ with $\mathbf{g}^{\text{MMSE}}$ and $\mathbf{u}^{\text{MMSE}}$\\
    5) Update $\mathbf{P}^{[n]}$ by solving \eqref{eWMMSE} under fixed $\mathbf{d}^{[n]}$,  $\mathbf{a}_{p}^{[n]}$\\
    6) Repeat steps 2 and 4 until convergence\\
    			 	 			\hline
    			 	 		\end{tabular}\vspace{-1.25em}
    			 	 	\end{table}

\section{Numerical results}
\begin{figure}[t!]
\centerline{\includegraphics[scale=0.47]{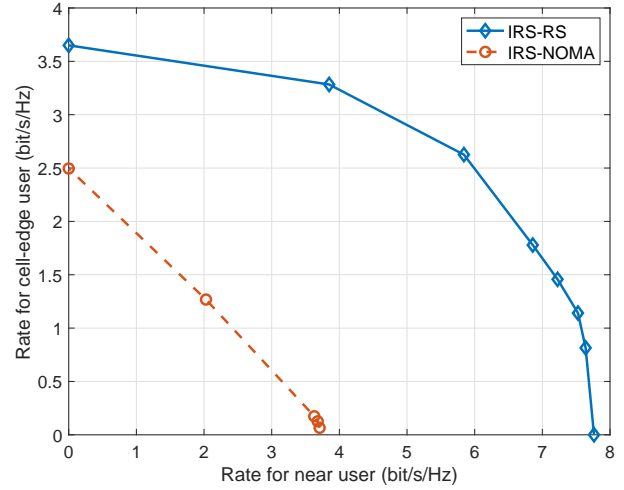}}\vspace{-1.0em}
\caption{Illustration of the rate region with perfect CSIT.}
\label{fig2}\vspace{-1.15em}
\end{figure}
In this section, we numerically investigate the performance of RS based IRS (hereinafter termed as IRS-RS) aided communication network. In particular, while we evaluate the proposed RS algorithm, we also compare it with the baseline IRS-NOMA from \cite{Ding2020}. Both perfect and imperfect CSIT are considered, whereby to model the imperfect CSIT, we consider the stochastic error model as given in~\cite{Aquilina2017}. While the level of tolerance (\emph{i.e.,} the MSE difference between two consecutive iterations) for the proposed alternating algorithm is set at $10^{-4}$, the maximum iteration number is fixed at 50. %
To explicitly compare the rate regions of IRS-RS with IRS-NOMA through two-dimensional figures, we consider a cell with 2 users, 1 near and 1 cell-edge. The number of antennas at the BS, $N_{T}=4$ and the number of reflecting elements at each IRS, $N=20$. Also, the channel strength of both the channels, the one from BS to IRS and IRS to the cell-edge user is considered to be $0.3$ times the channel strength of the channel from BS to the near-user. 
%
Unless otherwise stated, the SNR is considered to be $20$dB for the rate region calculations. The boundary of these rate region graphs are calculated by varying the weights assigned to users. In this work, the weight of the near cell user is considered to be $u_{1} = 1$ and the weights of cell-edge user is varied as $u_{1}^{'}$ = $10^{[-3,-1,-0.75,-0.5,-0.25,....,0.75,1,3]}$ for both perfect and imperfect CSIT cases.

We begin by showing the rate regions of both IRS-RS and IRS-NOMA under perfect CSIT in Fig. \ref{fig2}. It can be seen from the figure that the individual rates achieved by both the near and cell-edge users are greater for the case of IRS-RS than in IRS-NOMA. Further, the total rate region achieved by IRS-RS is greater than IRS-NOMA. 
\begin{figure}[t!]
\centerline{\includegraphics[scale=0.47]{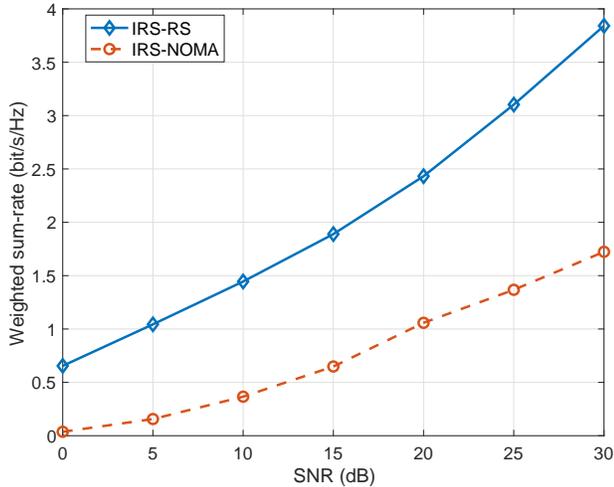}}
\caption{WSR vs SNR with perfect CSIT.}
\label{fig3}\vspace{-1.0em}
\end{figure}

Next, in Fig. \ref{fig3} the weighted sum-rate is plotted with respect to varying SNR values for perfect CSIT.  It can be observed that IRS-RS achieves higher sum-rate than IRS-NOMA across the entire SNR region. Further, the gap in performance increases with the increase in SNR.
%
%
\begin{figure}[t!]
\centerline{\includegraphics[scale=0.47]{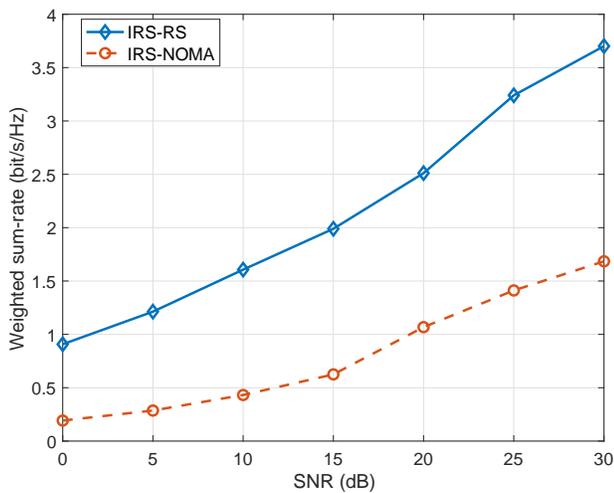}}
\caption{WSR vs SNR with imperfect CSIT.}
\label{fig5}\vspace{-1.0em}
\end{figure}

Finally, in Fig. \ref{fig5} the WSR vs SNR graphs for both IRS-RS and IRS-NOMA are shown under imperfect CSIT. Here, it is assumed that users are able to estimate the channel perfectly while the instantaneous channel estimated at the BS is imperfect. Similar to the case of perfect CSIT, it can be seen that IRS-RS outperforms IRS-NOMA significantly. 

\section{Conclusion}
A novel IRS-RS framework to improve the performance of a multi-user downnlink network was proposed, whereby RSMA was applied at the BS and IRSs with multiple reflecting elements were used to assist cell-edge users. A WSR optimization problem was formulated for the joint selection of reflection coefficients of IRS and design of beamformers at the BS subject to the constraints of transmission power at the BS, QoS at each user, and finite resolution for the elements of the IRS. To tackle the non-convexity of the original optimization problem, a modified WMMSE based algorithm was proposed to obtain the solution to the problem. A baseline NOMA-based transmission scheme for the considered IRS network was also implemented for the purpose of comparison. Through extensive numerical results, it was shown that the proposed beamformer design for IRS-RS achieves better performance when compared to the IRS-NOMA transmission scheme.

\end{document}